\documentclass[aps,onecolumn,notitlepage,preprintnumbers,superscriptaddress,amsmath,amsfont,amssymb,graphicx,10pt] {revtex4}
\usepackage{epsfig,amsmath,amssymb}
\usepackage{color}
\usepackage{float}
\definecolor{BrickRed}{rgb}{0.9,0.1,0}
\begin{document}
\preprint{DO-TH 15/03}
\title{Discerning new physics in charm meson leptonic and semileptonic decays }
\author{Svjetlana Fajfer}
 \email[Electronic
address:]{svjetlana.fajfer@ijs.si} 
\affiliation{Department of Physics, University of Ljubljana, Jadranska 19, 1000 Ljubljana, Slovenia}
\affiliation{J. Stefan Institute, Jamova 39, P. O. Box 3000, 1001
  Ljubljana, Slovenia}
  \author{Ivan Ni\v sand\v zi\' c}
 \email[Electronic address:]{ivan.nisandzic@tu-dortmund.de} 
\affiliation{Institut f\" ur Physik, Technische Universit\" at Dortmund, D-44221 Dortmund, Germany}
\author{Ur\v sa Rojec} 
\email[Electronic address:]{ursa.rojec@cosylab.com} 
\affiliation{Department of Physics, University of Ljubljana, Jadranska 19, 1000 Ljubljana, Slovenia}
\date{\today}

\begin{abstract}

Current experimental information on the charm meson decay observables in which the $c\to s\ell\nu_\ell$ transitions occur is well compatible with the Standard Model predictions. Recent precise lattice calculations of the $D_s$ meson decay constant and form factors in $D\to K\ell\nu$ decays offer a possibility to search for the small deviations from the Standard Model predictions in the next generation of the high intensity flavour experiments. We revisit constraints from these processes on the new physics contributions in the effective theory approach. We investigate new physics effects which might appear in the differential distributions for the longitudinally and transversely polarised $K^\ast$ in $D\to K^\ast\ell\nu_\ell$ decays. Present constraints from these observables are rather weak, but could be used to constrain new physics effects in the future. In the case of $D\to K\ell\nu$ we identify observables sensitive on new physics contribution coming from the scalar Wilson coefficient, namely the forward-backward and the transversal muon asymmetries. By allowing that new physics modifies only the second lepton generation but not the first one, we identify allowed region for the differential decay rate for the process $D\to K\mu\nu_\mu$ and find that it is allowed to deviate from the Standard Model prediction by only few percent. 
The lepton flavour universality violation can be tested in the ratio $R_{\mu/e}(q^2)\equiv \frac{d\Gamma^{(\mu)}}{dq^2}/\frac{d\Gamma^{(e)}}{dq^2}$.
If the first lepton generation behaves as in the Standard Model, we find, using current constraint on the scalar Wilson coefficient, that the ratio $R_{\mu/e}(q^2)$ is currently allowed to be within the range $(0.9, 1.2)$, depending on the value of $q^2$.
\end{abstract}

\maketitle
\section{Introduction}
After discovery of the Higgs boson, the main role of LHC became the search for particles which do not belong to the Standard Model (SM). The alternative way to investigate presence of physics beyond SM is to explore results from high precision experiments at low energies. Namely, very accurate theoretical handling of processes at low energies enables the extraction of constraints on possible new physics (NP) effects in these processes. At low energies, processes driven by flavour changing neutral currents were usually considered to be the best candidates to detect NP. However, the recent indications of the difference between experimental result for the branching fractions for $B \to D^{(\ast)} \tau\nu_\tau$ and the theoretical predictions (see e.g.~\cite{Fajfer:2012vx, Fajfer:2012jt}) open new window in searches for new physics at low energies in the processes induced by the charged currents.
The $c\to s \ell\nu_\ell$ transition within charm mesons 
might offer important tests of the SM and non-perturbative QCD dynamics in particular. In the past few years, the significant effort has been made in both theoretical and experimental research of these transitions. The precise value of the decay constant of $D_s$ meson is now known from the unquenched lattice QCD simulations that involve the effects of dynamical up, down, strange and charm quarks~\cite{Bazavov:2014wgs}. The shapes of the semileptonic form factors $f_{+,0}(q^2)$ for the process $D\to K\ell\nu$ over the whole physical $q^2$ region were also recently calculated in the lattice QCD~\cite{Koponen:2013tua}. On the experimental side, several new measurements of relevant branching fractions and the extraction of  form factors shapes have been performed.
The Belle Collaboration recently measured precisely the branching fractions of leptonic modes $D_s\to\ell\nu$, where $\ell=\mu,\tau$~\cite{Zupanc:2013byn}. 
The results of measurements of the branching fractions and the form factor shapes for the process $D\to K\ell\nu$ were reported by collaborations FOCUS, Belle, BaBar and CLEO~\cite{Link:2004dh, Widhalm:2006wz, Aubert:2007wg, Besson:2009uv, Ge:2008aa}. The analogous experimental results for the process $D\to K^\ast\ell\nu$ were presented in~\cite{Link:2005dp, Shepherd:2006tw, Briere:2010zc, delAmoSanchez:2010fd}. 

The theoretical predictions within the SM can be compared to the measured values of the total or differential branching fractions in order to extract the $\vert V_{cs}\vert$ element of Cabibbo-Kobayashi-Maskawa (CKM) matrix. On the other hand, the constrains on the effects of the new physics (NP) in a given process can be derived after fixing the value of the CKM matrix element from some independent source. In 2007, the $c\to s\ell\nu$ transitions attracted a lot of interest from the point of view of searches for NP, after the disagreement between the lattice evaluations of the decay constant $f_{D_s}$ and experimental extractions thereof at the level of around $4\sigma$. Several different NP scenarios were considered as explanations of that puzzle~\cite{Dobrescu:2008er, Dorsner:2009cu, Kronfeld:2008gu}. Current agreements between the experimental results and the lattice evaluations offer an opportunity for a derivation of tight constraints on the NP effects in these processes. The recent analysis of this kind was performed in Refs.~\cite{Barranco:2013tba, Barranco:2014bva}. These authors studied the leptonic $D_s \to \ell \nu_\ell$ and the semileptonic decays $D \to K \ell \nu_\ell$ within effective theory approach and using two specific models. In the present article we concentrate mainly on the non-standard (pseudo)scalar operators and include the discussion of the observables in the decays $D\to V\ell\nu$, $V=K^\ast,\phi$. In Sec.~II we introduce the effective Lagrangian. Sec.~III is devoted to constraints on the Wilson coefficient of the pseudoscalar operator coming from leptonic and semileptonic $D\to K^\ast\ell\nu$ decay mode. 
Sec.~IV contains the analyses of constraints on the Wilson coefficient arising from the scalar operator coming from $D\to K\ell\nu$. The branching ratio, differential branching ratio, the forward-backward and the transversal muon asymmetries in this process are considered. Sec.~V contains brief study of the right handed current, and the conclusions are given in Sec.~VI.
\section{The effective Lagrangian describing NP in $c\to s\ell\nu_\ell$ transitions}
We assume that the relevant NP states are considerably heavier than the typical hadronic energy scale so that they can be integrated out, together with the W boson, leading to the appearance of non-standard higher dimensional operators in the low energy effective description of $c\to s\ell\nu_\ell$ transitions. We choose the following normalisation of the effective Lagrangian:
\begin{equation}
\mathcal{L}_{eff}=-\frac{4 G_F}{\sqrt{2}}V_{cs}\sum_{\ell=e,\mu,\tau}\sum_{i} c_i^{(\ell)}\mathcal{O}_{i}^{(\ell)}+\mathrm{H.c.}.
\label{effective lagrangian}
\end{equation}
The usual four-fermion operator is $\mathcal{O}_{SM}^{(\ell)}=\big(\bar{s}\gamma_\mu P_L c\big)\big(\bar{\nu}_\ell\gamma^\mu P_L\ell\big)$ with the coefficient $c_{SM}^{(\ell)}=1$.
In this article we concentrate on the non-standard effective operators that involve the (pseudo)scalar quark and lepton densities, while keeping only the left-handed neutrinos, namely:
\begin{equation}
\mathcal{O}_{L(R)}^{(\ell)}=\big(\bar{s}P_{L(R)} c\big)\big(\bar{\nu}_\ell P_R\ell\big).\label{scalar operator}
\end{equation}
These operators might be induced by integrating out the beyond the SM charged scalar boson at the tree level. Such a boson can arise in a two-Higgs doublet model (THDM), the extension of the SM with an additional scalar doublet, c.f. the review article~\cite{Branco:2011iw}. The most studied such model is the so called type-II THDM, in which $c^{(\ell)}_{R(L)}$ can be expressed as the combination of the two real parameters: mass of the charged scalar $m_{H^+}$ and $\tan\beta$, the ratio of the vacuum expectation values of the two doublets. Since it has small number of free parameters, this model is actually tightly constrained by the flavour phenomenology and the new LHC results~\cite{Grinstein:2013npa, Eberhardt:2013uba}.
For generality, we allow the coefficients $c^{(\ell)}_{S,R(L)}$ to be complex valued and to depend on the flavour of the charged lepton. For example, additional dependence (besides the factor of $m_\ell$) on the charged lepton's flavour is present in the THDM of the type-III~\cite{Crivellin:2013wna}, originating from the non-holomorphic Yukawa couplings in the fermion mass basis. Another possibility is given by the aligned THDM~\cite{Pich:2009sp, Jung:2010ik} in which the Yukawa couplings of the fermions to the neutral scalars are flavour diagonal in the fermion mass basis, while the new sources of the CP violation stem from the complex Yukawa couplings involving the charged scalar.
In the following sections we constrain the values of the scalar Wilson coefficients of the operators in~\eqref{scalar operator} from the available measured values of the corresponding branching fractions of the (semi)leptonic decays.

It is also possible that the higher dimensional operators modify the $W\bar{s}c$ coupling, which would be  reflected in the low energy Lagrangian by the appearance of the non-standard admixture of the right-handed quark current,
\begin{equation}
\mathcal{O}_{V,R}^{(\ell)}=\big(\bar{s}\gamma_\mu P_R c\big)\big(\bar{\nu}_\ell\gamma^\mu P_L\ell\big).\label{Right handed}
\end{equation}
We briefly study the effects of this operator in section~\ref{Right handed section}.
The tensor operator $\big(\bar{s}\sigma_{\mu\nu}P_{R}c\big)\big(\bar{\nu}_\ell\sigma^{\mu\nu} P_R\ell\big)$ could also appear~\cite{Barranco:2013tba, Barranco:2014bva}, together with the (pseudo)scalar operators, after integrating out a scalar leptoquark at the tree level. We ignore these contributions for the present lack of reliable information of the tensor form factors. 
\section{The Wilson coefficient of the pseudo-scalar operator} 
\subsection{NP in $D_s\to\ell\nu_\ell$}
In this section we derive the constraints on the linear combination of the Wilson coefficients $c_{L(R)}^{(\ell)}$ from the measured branching fractions of the purely leptonic $D_s\to\ell\nu$ decay modes. The hadronic matrix element of the corresponding axial vector current is parametrized by the decay constant $f_{D_s}$ via $\langle 0\vert\bar{s}\gamma_\mu\gamma_5\vert D_s(k)\rangle=f_{D_s}\,k_\mu$. Using the identity $\partial_\mu (\bar{s}\gamma_\mu\gamma_5 c)=i\,(m_s+m_c)\bar{s}\gamma_5 c$ one finds that the $f_{D_s}$ suffices to parametrize the matrix element of the pseudoscalar density, 
\begin{equation}
\langle 0\vert \bar{s}\gamma_5 c\vert D_s(k)\rangle=\frac{f_{D_s} m_{D_s}^2}{m_c+m_s}.\label{PCAC1}
\end{equation}
The standard formula for the branching fraction is then modified to the following form
\begin{equation}
\mathcal{B}(D_s\to\ell\nu_\ell)=\tau_{Ds}\frac{m_{Ds}}{8\pi}f_{Ds}^2\Bigg(1-\frac{m_\ell^2}{m_{Ds}^2}\Bigg)^2G_F^2(1+\delta_{em}^{(\ell)})\vert V_{cs}\vert^2m_\ell^2\Bigg\vert 1-c_P^{(\ell)}\frac{m_{Ds}^2}{(m_c+m_s)m_\ell}\Bigg\vert^2,
\end{equation}
where the pseudoscalar combination of the couplings is $c_P^{(\ell)}\equiv c_{R}^{(\ell)}-c_{L}^{(\ell)}$.
In the evaluation of the constraints we use the latest theoretical value of the decay constant $f_{D_s}=249.0(0.3)(^{+1.1}_{-1.5})\,$MeV, calculated in the lattice QCD with sub-percent precision by the Fermilab Lattice and MILC collaborations~\cite{Bazavov:2014wgs}. At this level of precision it is mandatory to take into account the uncertainty in the lifetime of $D_s$ meson ($1.4\%$) and the electromagnetic corrections parametrized by $\delta_{em}^{(\ell)}$. The detailed study of the electromagnetic effects is out of scope of the present article; we draw attention to Ref.~\cite{Becirevic:2009aq} for detailed analysis regarding the $B\to \ell\nu\gamma$ process and comparison with the $D\to\ell\nu\gamma$ case. There are several contributions to $\delta_{em}^{(\ell)}$: the long distance soft photon corrections that can be studied in the approximation of point-like charged mesons and leptons, the universal short distance electroweak corrections, and the contributions that probe the hadronic structure of the process and require the knowledge of additional hadronic form factors. Following~\cite{Bazavov:2014wgs, Becirevic:2009aq} we estimate the $\delta_{em}^{(\mu)}$ to be in the range $\sim (1-3)\%$ and $\delta_{em}^{(\tau)}\sim (0-1)\%$, and include these values as the new sources of the uncertainty.
\begin{figure}[H]
\begin{center}
\includegraphics[width=0.30\textwidth]{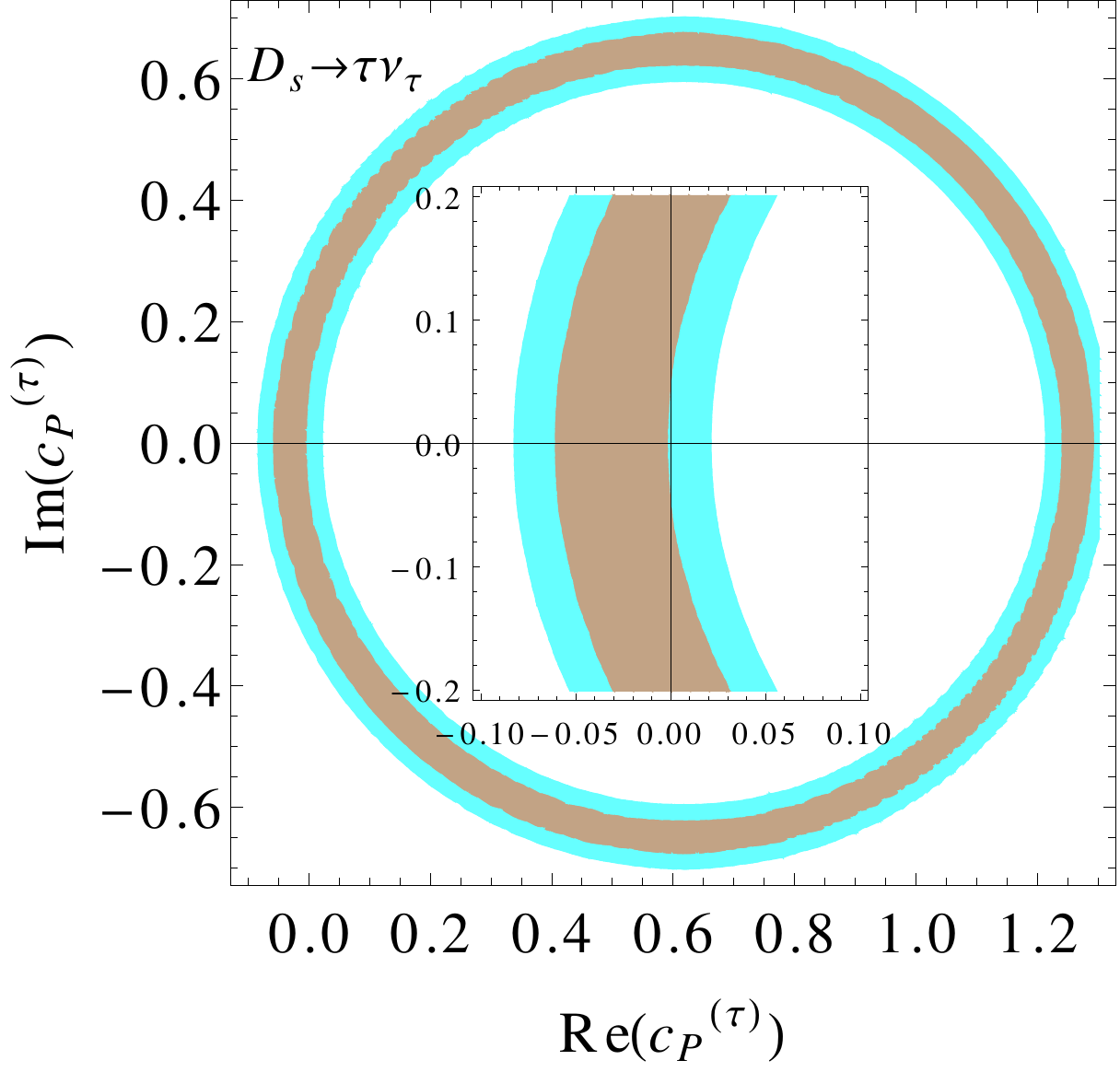}
\includegraphics[width=0.315\textwidth]{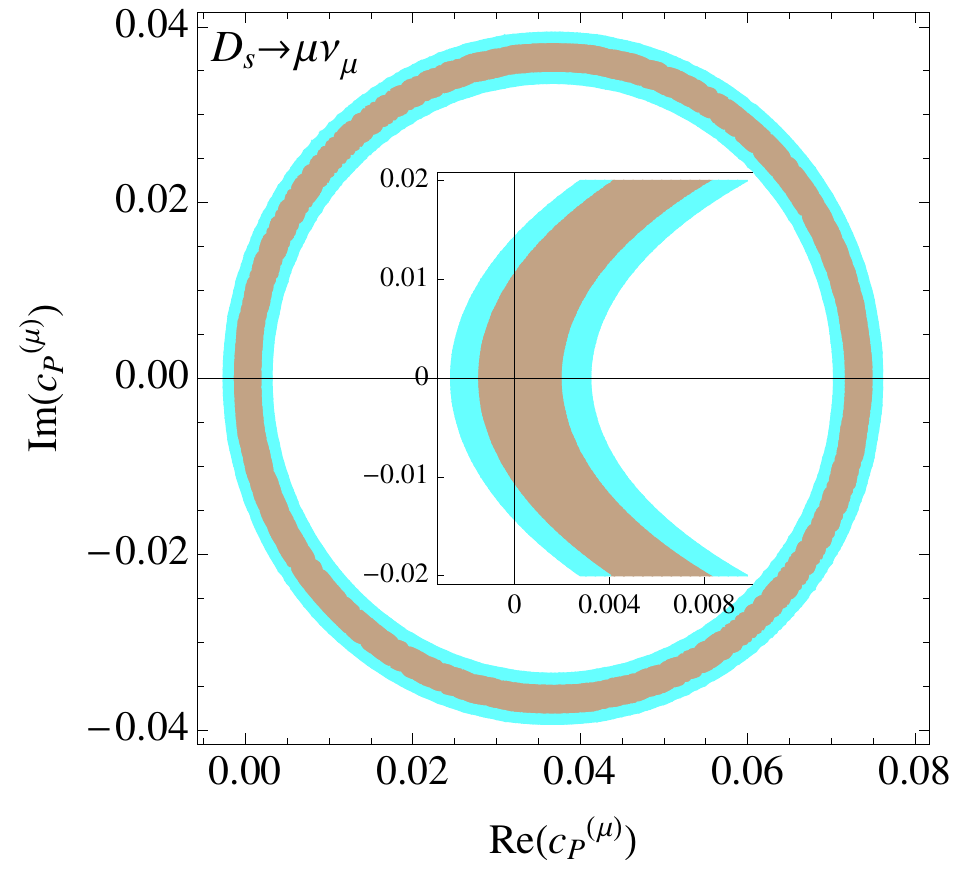}
\end{center}
\caption{Allowed regions of the effective coupling $c_P^{(\tau)}$ (left panel) and $c_P^{(\mu)}$ (right panel), extracted from the branching fraction of the decay mode $D_s\to\tau(\mu)\nu$, respectively. The $68\%$ ($95\%$) C.L. regions of the parameters are shown in darker (lighter) shades.}\label{Fig.1a}
\end{figure}
The leptonic branching fractions of $D_s^+\to\tau^+(\mu^+)\nu$ were recently measured by the Belle Collaboration~\cite{Zupanc:2013byn}. The measured values, together with the upper limit of yet unobserved channel $D_s^+\to e^+\nu$ were given as following:
\begin{eqnarray}
\label{experimental results}
\mathcal{B}(D_s\rightarrow \ell\nu_\ell)=
\begin{cases}
(5.7\pm 0.21^{+0.31}_{-0.3})\%,	&  D_s \rightarrow \tau\nu_\tau,\\
(0.531\pm 0.028\pm 0.020)\%,		& D_s \rightarrow  \mu \nu_\mu,\\
< 1.0\cdot 10^{-4}\,,\,\text{95\%\,C.L.},	& D_s \rightarrow e\nu_e. \label{Exp-leptonic}
\end{cases}
\end{eqnarray}
We use the value of the $\vert V_{cs}\vert$ which results from the global fit of the unitary 
CKM matrix and given by the CKMFitter Collaboration~\cite{Charles:2004jd}, $V_{cs}=0.97317^{+0.00053}_{-0.00059}$, for we do not expect this value to be influenced by the operators in~\eqref{scalar operator}. The resulting allowed parameter space of the corresponding NP couplings is visualized in the Fig.~\ref{Fig.1a}. The upper limit in~\eqref{Exp-leptonic} leads to the constraint~$\vert c_P^{(e)}\vert<0.005$. 

One could also consider the ratios of the branching fractions, i.e.~$R_{\tau/\mu}=\mathcal{B}(D_s\to\tau\nu)/\mathcal{B}(D_s\to\mu\nu)$ as a test of the lepton flavour universality of the charged current. This quantity has small theoretical error that comes from the uncertainties in masses of the particles involved in the process, see e.g.~\cite{Zupanc:2013byn}. It stays unchanged with respect to SM in the natural flavour conserving THDMs, but could receive corrections e.g. in the type-III THDM from the non-holomorphic Yukawa couplings in the fermion mass basis. Careful investigation of this ratio should also include the effects of the electromagnetic corrections. 
\subsection{NP in $D\to K^\ast\ell\nu_\ell$}
The pseudoscalar Wilson coefficient $c_P^{(\ell)}$ contributes also to the semileptonic decays of the pseudoscalar to vector mesons. These processes offer larger number of observables than the two-body leptonic decays due to the existence of the non-trivial angular distributions, see e.g.~\cite{Fajfer:2012vx}. The information about the helicity suppressed contribution can be extracted experimentally by comparing the decays that involve electrons and muons in the final state. This is, however, difficult task at present but could be performed more precisely in the next generation of flavour experiments~\cite{Aushev:2010bq, Asner:2008nq}.
The helicity suppressed contributions are also sub dominant, which implies that the sensitivity of the processes $D\to K^\ast\ell\nu_\ell$ and $D_s\to \phi\ell\nu_\ell$ to the coefficient $c_P^{(\ell)}$ is weaker when compared to the pure leptonic decays. Also, the knowledge of the form factors in these transitions is currently less precise. The information about the decay mode $D\to K^\ast\ell\nu$ is reconstructed from the experimentally observed $D\to K\pi\ell\nu$ process in which the dominant vector intermediate state interferes with scalar $K\pi$ amplitude and also, to smaller extent, with higher waves~\cite{Link:2002ev}. The extraction of the possible NP effects from the angular analysis thus requires careful disentangling of such resonant (and also other non-resonant) contributions. The lattice simulations provide easier access to the form factors for the process $D_s\to\phi\ell\nu$, in which none of the two mesons contains the light valence quarks and the $\phi$ meson can be treated as stable to a good approximation. The first results of such a calculation (including the scalar form factor $A_0(q^2)$, to be defined below) were recently presented by the HPQCD Collaboration~\cite{Donald:2013pea}.

The standard parametrization of the hadronic matrix element of the vector and axial vector currents in terms of form factors $V(q^2)$ and $A_{0,1,2}(q^2)$ is as in~\cite{Wirbel:1985ji}:
\begin{equation}
\begin{split}
\langle V(k',\epsilon)\vert\bar{s}\gamma_\mu(1-\gamma_5) c\vert P(k)\rangle &=\epsilon_{\mu\nu\alpha\beta}\;\frac{2 i\,V(q^2)}{m_P+m_V}
\epsilon^{*\nu}k^\alpha k^{'\beta}
-(m_P+m_V)\bigg(\epsilon_\mu-\frac{\epsilon\cdot q q^\mu}{q^2}\bigg)A_1(q^2)+\\
&+\epsilon\cdot q\bigg(\frac{(k+k')_\mu}{m_P+m_V}-\frac{m_P-m_V}{q^2}q_\mu\bigg)A_2(q^2)-2\,m_{V}\frac{\epsilon\cdot q q^\mu}{q^2}A_0(q^2),
\label{Hadronic}
\end{split}
\end{equation}
\begin{equation}
A_3(q^2)\equiv\frac{m_P+m_{V}}{2m_{V}}A_1(q^2)-\frac{m_P-m_{V}}{2m_{V}}A_2(q^2),
\end{equation}
where the spurious singularity at $q^2=0$ is avoided with the constraint $A_3(0)=A_0(0)$.
In the above formulas the four-vector $\epsilon$ denotes the polarization vectors of a spin-1 meson, while the transferred four-momentum is $q\equiv k-k'=p_\ell+p_\nu$. 
Contracting the above matrix element with $q^\mu$ one derives the parametrization of the pseudoscalar density in terms of form factor $A_0(q^2)$,
\begin{equation}
\langle V\vert \bar{s}\gamma_5 c\vert P\rangle=\frac{2m_V \epsilon^\ast\cdot q}{m_c+m_s}A_0(q^2).
\end{equation}
The differential decay rates of the process can be conveniently expressed in terms of hadronic helicity amplitudes that are defined as projections of the the matrix element of the hadronic current \eqref{Hadronic} to the polarization vectors of the charged lepton-neutrino pair $\tilde\epsilon_m^\mu$, where $m$ denotes the polarizations $t,0,\pm$. These amplitudes are explicitly given in Appendix~\ref{Appendix1}. Note that only the helicity amplitude $H_{t}(q^2)$, which receives the contribution from terms with $A_0(q^2)$, is modified in the presence the pseudoscalar Wilson coefficients,
\begin{equation}
H_{t}\rightarrow \bigg(1-c_P^{(\ell)}\frac{q^2}{m_\ell(m_c+m_s)}\bigg)H_{t}.
\end{equation}
The form factors are analytic functions of $q^2$ in the physical region and satisfy the dispersion relations by the conditions of causality and unitarity. Most of the experimental measurements of the form factors assume the single pole dominance behaviour by which the main contribution in the dispersion relations arises from the lowest pole outside the physically allowed region. In the Ref.~\cite{Fajfer:2005ug} the form factors for $D\to K^\ast\ell\nu$ transitions were studied in the framework that combines the heavy quark and chiral symmetries and includes the effects of the resonances beyond the simple pole approximation. The authors of~\cite{Melikhov:2000yu} employ the dispersion approach within the constituent quark model. In 2005 the FOCUS Collaboration performed the non-parametric measurements of the hadronic helicity amplitudes~\cite{Link:2005dp} as functions of the lepton pair invariant mass in several bins. However, the errors in this study are too large to be used in constraining NP contributions. The latest analysis of the $D\to K\pi\ell\nu$ decays was performed by the BaBar~\cite{delAmoSanchez:2010fd}. They used the simple pole parametrization of form factors and extracted the values of the ratios of the form factors for the $D\to K^\ast$ transition at the single kinematic point: $V(0)/A_1(0)=1.463\pm 0.035$, $A_2(0)/A_1(0)=0.801\pm 0.03$, $A_1(0)=0.6200\pm 0.0057$. Since only electrons and positrons were used, the analysis remained insensitive to the form factor $A_0(q^2)$. 

\begin{figure}[H]
\begin{center}
\includegraphics[width=5.5cm,clip=true]{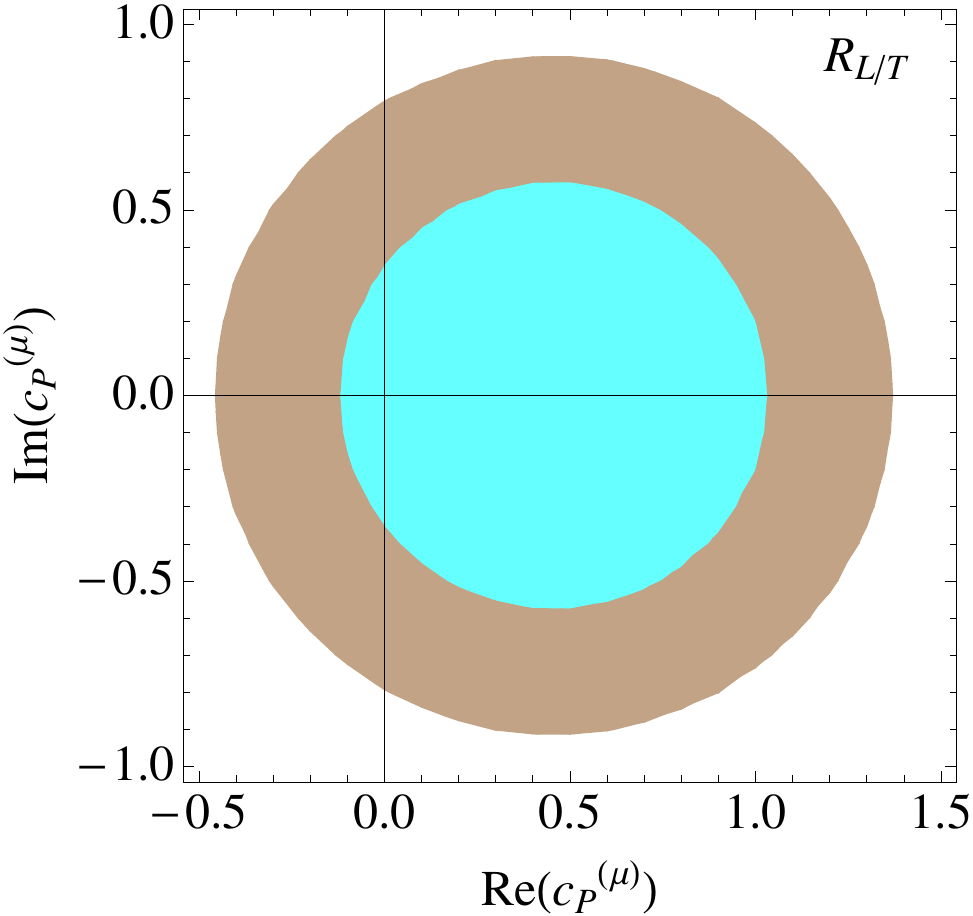}
\end{center}
\caption{Allowed regions of the effective coupling $c_P^{(\mu)}$, extracted from the ratio $R_{L/T}$. The colour coding follows the convention of Fig.\ref{Fig.1a}.}\label{Fig.Xa}
\end{figure}

In order to get an estimate of the allowed NP contributions in $D\to K^\ast\ell\nu$ we proceed by using the constraint $A_3(0)=A_0(0)$ to infer the value of $A_0(0)$ and assume that the dependence on the $q^2$ of the form factor $A_0(q^2)$ is as well described with the simple pole parametrization. We then consider $R_{L/T}$, the ratio of the decay widths of the longitudinally and transversally polarized $K^\ast$ fractions, as an observable which is sensitive to $c_P^{(\ell)}$. The differential distributions for the longitudinally and transversely polarized $K^\ast$ are:
\begin{equation}
\begin{split}
\frac{d\Gamma_L}{dq^2}&=\mathcal{N}(q^2)\bigg(1-\frac{m_\ell^2}{q^2}\bigg)^2\bigg[\big(1+\frac{m_\ell^2}{2q^2}\big)\vert H_0\vert^2+\frac{3m_\ell^2}{2q^2}\vert H_t\vert^2\bigg],\quad
\frac{d\Gamma_T}{dq^2}=\mathcal{N}(q^2)\bigg(1-\frac{m_\ell^2}{q^2}\bigg)^2\bigg[\big(1+\frac{m_\ell^2}{2q^2}\big)\big(\vert H_+\vert^2+\vert H_-\vert^2\big)\bigg],
\end{split}
\end{equation}
where the overall factor is $\mathcal{N}(q^2)=G_F^2\vert V_{cs}\vert^2 q^2\vert\mathbf{q}\vert/(96\pi^3m_D^2)$. We use the Particle Data Group (PDG) averaged value~\cite{Agashe:2014kda} of the ratio $R_{L/T}=1.13\pm 0.08$ for the process $D^+\to\bar{K}^{\ast0}\mu^+\nu$ to extract the allowed regions of the coefficient $c_P^{(\mu)}$ coupling in Fig.~\ref{Fig.Xa}.
The resulting constraint turns out to be currently much	 weaker than the one shown in Fig.~\ref{Fig.1a}. 
It is expected that the collaborations Belle II~\cite{Asner:2008nq} and BESIII~\cite{Aushev:2010bq} are going to measure the processes $D_{(s)}\to K^\ast(\phi)\ell\nu$ with an enhanced precision. 
Given the possible lattice QCD improvements, these processes could serve as the useful complementary source of information about the NP in $c\to s\ell\nu$ transitions in the near future.

\section{The Wilson coefficient of the scalar operator} 

The semileptonic $D\to K\ell\nu$ decays are affected by the scalar combination of the Wilson coefficients $c_S^{(\ell)}=c_R^{(\ell)}+c_L^{(\ell)}$. We use the latest lattice evaluation of the corresponding form factors and measured values of the branching fractions to constraint the values of $c_S^{(\ell)}$, $\ell=e,\mu$. Then we introduce the forward-backward and the transversal muon asymmetries as the observables that can be used to extract further constraints on the real and imaginary parts of the scalar Wilson coefficient, respectively.
\subsection{NP from branching fractions $\mathcal{B}(D\to K\ell\nu_\ell)$}
The hadronic matrix element of the vector current for the $D(k)\to K(k')\ell\nu_\ell$ decay is parametrized by form factors $f_{+,0}(q^2)$ as
\begin{equation}
\langle K(k')\vert \bar s \gamma_\mu c\vert D(k)\rangle = f_+ (q^2) \bigg((k +k')_\mu-\frac{m_D^2-m_K^2}{q^2} q_\mu\bigg)+f_0 (q^2) \frac{m_D^2-m_K^2}{q^2} q_\mu\,,
\label{hadronic current}
\end{equation}
with the usual kinematic constraint $f_+(0)=f_0(0)$. The partially conserved vector current identity, $\partial_\mu (\bar{s}\gamma_\mu c)=i\,(m_s-m_c)(\bar{s}c)$, relates the matrix element of the scalar density to the form factor $f_0(q^2)$:
\begin{equation}
\langle K\vert \bar s c\vert D\rangle=\frac{m_D^2-m_K^2}{m_s-m_c}f_0(q^2).
\end{equation}
Non-vanishing hadronic helicity amplitudes for the transition $D\to K\ell\nu$ are $h_{0,t}=\tilde \epsilon^{\mu\ast}_{0,t}\langle K\vert J_\mu\vert D\rangle$ and are given explicitly by:
\begin{equation}
\begin{split}
&h_0(q^2)=\frac{\sqrt{\lambda(m_D^2,m_K^2,q^2)}}{\sqrt{q^2}}f_+(q^2),\quad h_t(q^2)=\Bigg(1+g_S^{(\ell)}\frac{q^2}{m_\ell(m_s-m_c)}\Bigg)\frac{m_D^2-m_K^2}{\sqrt{q^2}}f_0(q^2)\label{eq:1},
\end{split}
\end{equation}
where $\lambda$ denotes the function $\lambda(x,y,z)=x^2+y^2+z^2-2(xy+xz+yz)$. The formula for the differential decay rate of the process $D\to K\ell\nu_\ell$ is given by the formula
\begin{equation}
\frac{d\Gamma^{(\ell)}}{dq^2}=\frac{G_F^2\vert V_{cs}\vert^2\vert \mathbf{q}\vert q^2}{96\pi^3 m_D^2}\Bigg(1-\frac{m_\ell^2}{q^2}\Bigg)^2\Bigg[\vert h_0(q^2)\vert^2\Bigg(1+\frac{m_\ell^2}{2q^2}\Bigg)+\frac{3m_\ell^2}{2q^2}\vert h_t(q^2)\vert^2\Bigg], \label{decay-rate}
\end{equation}
where $\vert \mathbf{q}\vert=\sqrt{\lambda(m_D^2,m_K^2,q^2)}/2m_D$ is the magnitude of the transferred three-momentum in the rest frame of D meson. 
\begin{figure}[H]
\begin{center}
\includegraphics[width=5.5cm]{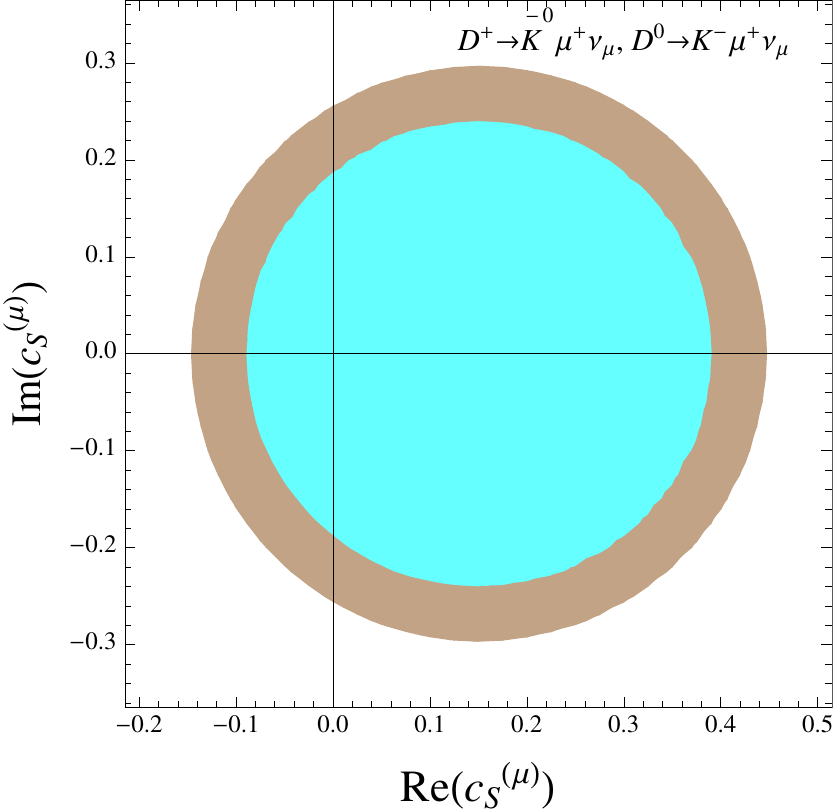}\label{Figure3}
\caption{Allowed regions of the effective coupling $c_S^{(\mu)}=c_R^{(\mu)}+c_L^{(\mu)}$ extracted from the branching fraction of the decay mode $D\to K\mu^+\nu$. The colour coding is the same as in the Fig.~\ref{Fig.1a}}
\end{center}
\end{figure}
The current average values of the branching fractions of the $D\to K\ell\nu_\ell$ decays can be found in the PDG review~\cite{Agashe:2014kda}:
\begin{eqnarray}
\label{experimental results}
\mathcal{B}(D\to K\ell\nu_\ell)=
\begin{cases}
(8.83\pm 0.22)\%, &  D^+ \rightarrow \bar K^0 e^+ \nu_e,\\
(9.2\pm 0.6)\%	,	& D^+ \rightarrow  \bar K^0 \mu^+ \nu_\mu, \\
(3.55\pm 0.04)\%,	& D^0 \rightarrow K^- e^+ \nu_e,\\
(3.30\pm 0.13)\%,	& D^0 \rightarrow K^-\mu^+ \nu_\mu.\end{cases}
\end{eqnarray}
The functional dependence on the $q^2$ of the form factors $f_{+,0}$ was recently calculated in lattice QCD by the HPQCD collaboration in Ref.~\cite{Koponen:2013tua}. 
Using their results and the measured branching fractions~\eqref{experimental results} we derive the constraint on the Wilson coefficients $c_S^{(\mu)}\equiv c_R^{(\mu)}+c_L^{(\mu)}$ and represent it in Fig.~3. In the case of electron, the $95\%$ C.L. interval reads: $\vert c_S^{(e)}\vert< 0.2$. The CLEO collaboration measured~\cite{Ge:2008aa} the differential decay rate for the process with electrons in the final state. The corresponding constraint is not significantly more stringent than the one obtained from the full branching ratio, see~\cite{Barranco:2013tba, Barranco:2014bva}. In Fig.~\ref{Fig.5} we present the sensitivity of yet unmeasured differential decay rate $d\Gamma^{(\mu)}/dq^2$ to the presently allowed values of the coupling $c_S^{(\mu)}$. We derive the allowed range for the ratio $R_{\mu/e}(q^2)\equiv \frac{d\Gamma^{(\mu)}}{dq^2}/\frac{d\Gamma^{(e)}}{dq^2}$ assuming $c_S^{(e)}=0$ and visualize it in the right panel of Fig.~\ref{Fig.5}. In the future precision measurements of the Belle II and at the high intensity tau-charm factories this ratio might serve as an excellent test of the lepton flavour universality. 
\begin{figure}[H]
\begin{center}
\includegraphics[width=7.1cm,clip=true]{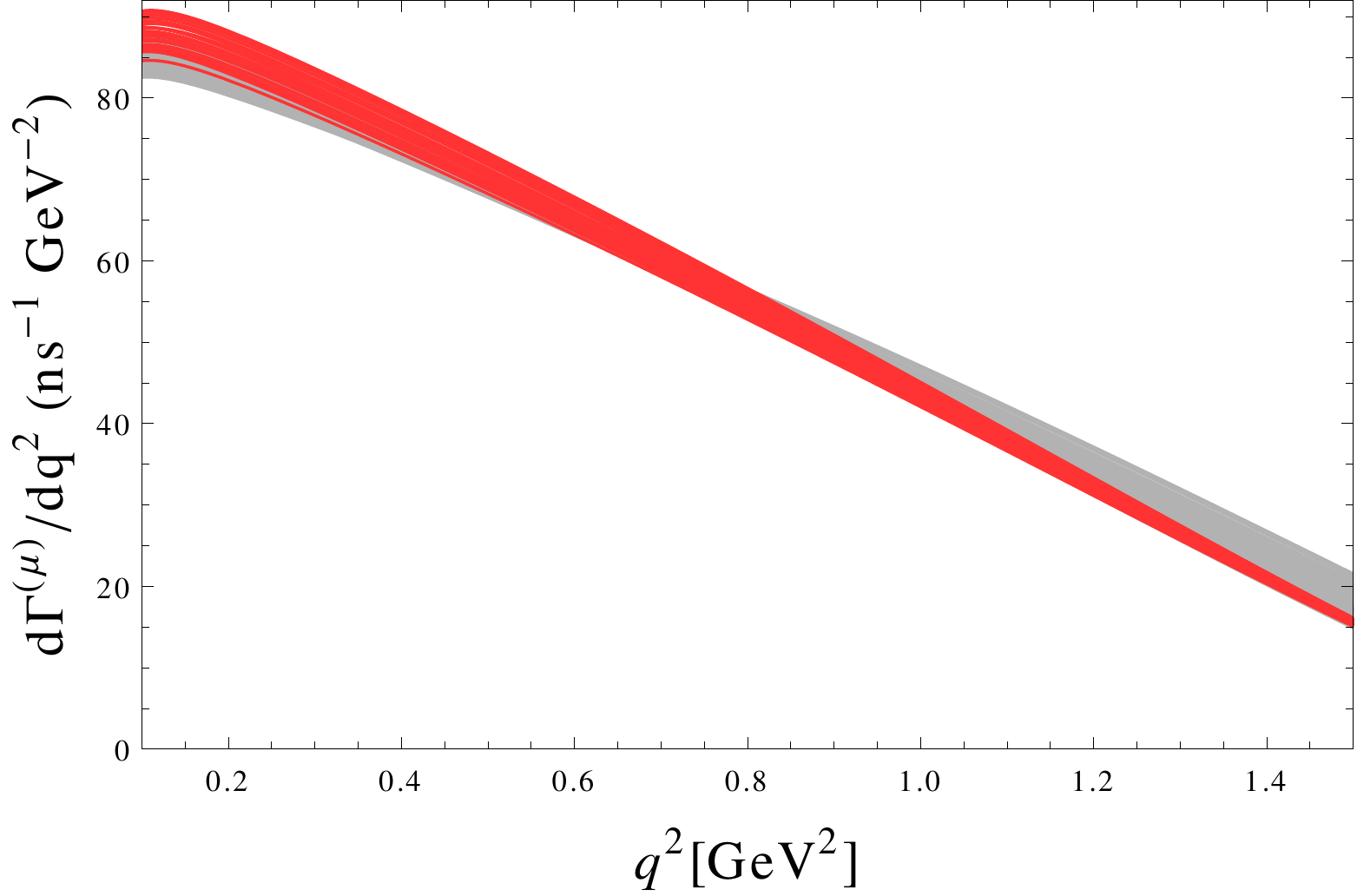}
\includegraphics[width=7.1cm,clip=true]{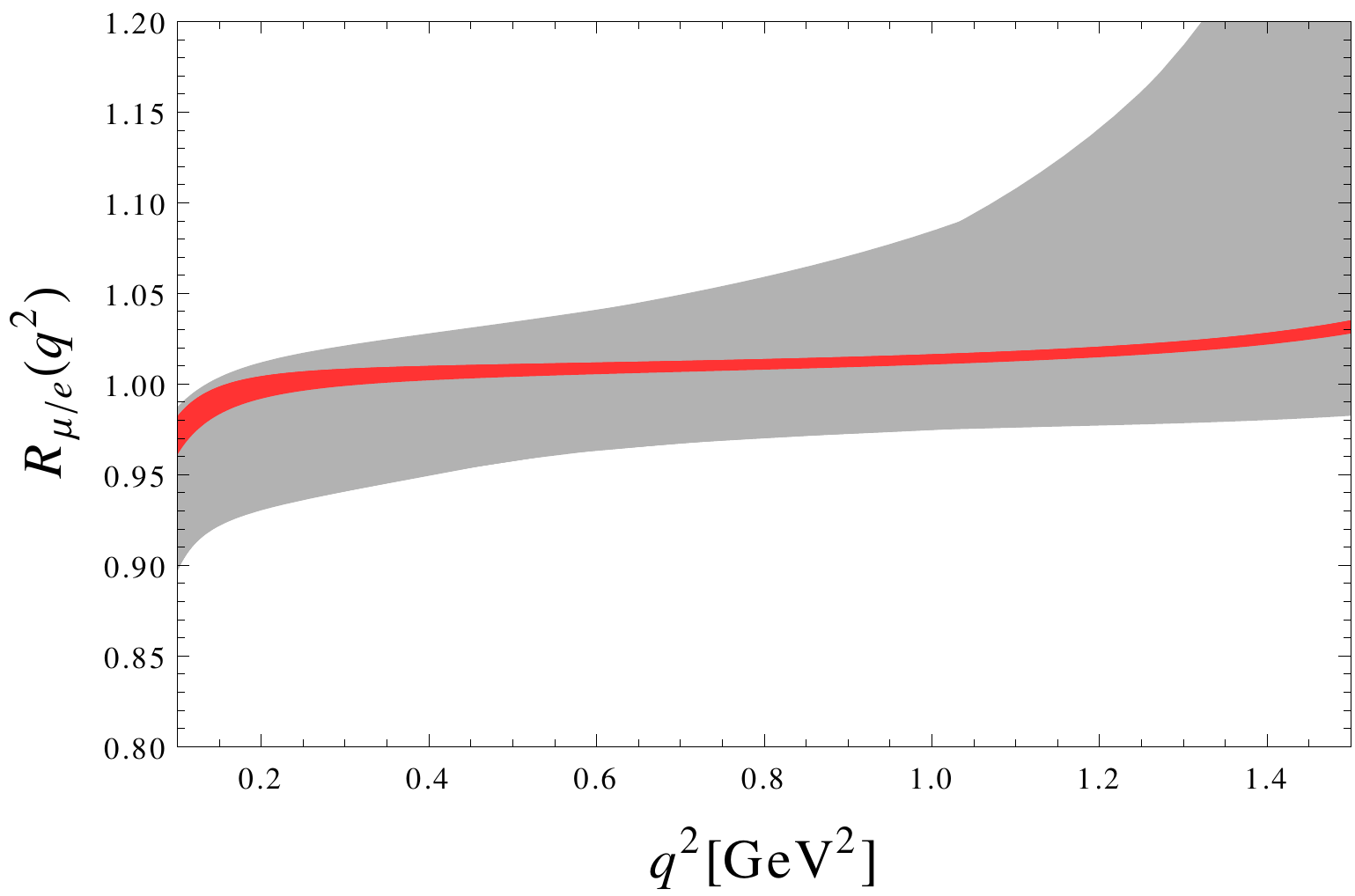}
\end{center}
\caption{Left panel: The differential decay rate for the process $D\to K\mu\nu_\mu$. The thin red band shows the SM prediction, while its width represents the uncertainty. The (wider) grey band corresponds to the deviations that result from the presently allowed scalar Wilson coefficient from the Fig.3. Right panel: the SM prediction and allowed deviations in the ratio $R_{\mu/e}(q^2)\equiv \frac{d\Gamma^{(\mu)}}{dq^2}/\frac{d\Gamma^{(e)}}{dq^2}$ assuming $c_S^{(e)}=0.$} \label{Fig.5}
\end{figure}

\subsection{NP in forward-backward asymmetry in $D\to K\ell\nu_\ell$}
It is instructive to introduce the observables which are exclusively sensitive to the real or imaginary parts of the Wilson coefficients. We first consider the differential decay distribution over the $\cos\theta_\ell$, where the $\theta_\ell$ is defined as the angle between the three-momenta of the $K$ meson and the charged lepton in the rest frame of the lepton-neutrino pair,
\begin{equation}
\frac{d^2\Gamma^{(\ell)}}{dq^2d\cos\theta_\ell}=a_\ell(q^2)+b_\ell(q^2)\cos\theta_\ell+c_\ell(q^2)\cos^2\theta_\ell.
\end{equation}
Note that the information carried by the function $b_\ell(q^2)$ is lost after integrating the above distribution over the angle $\theta_\ell$. This information can be accessed by measuring the forward-backward asymmetry in the angle $\theta_\ell$, defined as following:
\begin{equation}
\begin{split}
A_{FB}^{(\ell)}(q^2)\equiv \frac{\int_{-1}^0\frac{d^2\Gamma^{(\ell)}(q^2)}{dq^2d\cos\theta_\ell}d\cos\theta_\ell-\int_{0}^1\frac{d^2\Gamma^{(\ell)}(q^2)}{dq^2d\cos\theta_\ell}d\cos\theta_\ell}{d\Gamma^{(\ell)}/dq^2(q^2)}=
-\frac{b_\ell(q^2)}{d\Gamma^{(\ell)}(q^2)/dq^2}.\label{forward backward K}
\end{split}
\end{equation}
The above ratio has a small theoretical error in the full $q^2$ region due to the precise evaluation of the form factors and partly due to the cancellation of the uncertainties in the numerator and the denominator.
The function $b_\ell(q^2)$, given by
\begin{equation}
b_\ell(q^2)=-\frac{G_F^2\vert V_{cs}\vert^2\vert \mathbf{q}\vert q^2}{128\pi^3m_D^2}\Bigg(1-\frac{m_\ell^2}{q^2}\Bigg)^2\frac{m_\ell^2}{q^2}\,2Re(h_0h^{\ast}_t),
\end{equation}
is linearly sensitive to the real part of the coupling $c_S^{(\ell)}$. 
We illustrate the possible effects of the scalar operator on the forward-backward asymmetry in Fig.~5, with the values of $c_S^{(\mu)}$ taken from the $68\%$ C.L. allowed region in Fig.~3. The thin coloured (red) band represent the hadronic uncertainty in the shape of this function in the SM. The larger coloured band (grey) represents the currently allowed deviations from the SM. We conclude that the large deviations from the SM in this observable are not excluded at the present. The quantity $\mathcal{A}_{FB}^{(e)}$ is highly suppressed and insensitive to the corresponding scalar Wilson coefficient due to the tiny mass of the electron. The average value of the forward-backward asymmetry, $\langle {A}_{FB}^{(\ell)}\rangle$, can be calculated by performing the integration over the $q^2$ in the numerator and denominator of Eq.~\eqref{forward backward K}.
The SM value is $\langle A_{FB}^{(\mu)}\rangle =0.055(2)$. For various values of $c_S^{(\mu)}$ from the $68\%$ C.L. region in Fig.~3 this quantity can have values in the interval $(0,0.065)$.

Some comments about the NP scenarios that could affect these observables are in order here.
In the type-II THDM the Wilson coefficients that contribute to the $c\to s\ell\nu_\ell$ transitions are small:
\begin{equation}
c_{L}^{(\ell)}=\frac{m_s m_\ell\tan^2\beta}{m_{H^+}^2},\quad c_{R}^{(\ell)}=\frac{m_c m_\ell}{m_{H^+}^2},
\end{equation}
implying $c_S^{(\ell)}\simeq -c_P^{(\ell)}=c_{L}^{(\ell)}$. The values of the scalar and pseudoscalar couplings are thus approximately related, so that the tight constraints from the leptonic decays imply that the forward-backward asymmetry in $D\to K\mu\nu$ would not show the deviations from the SM. In more general THDMs the scalar and pseudoscalar coefficients are independent. Examples of such models are the Aligned THDM~\cite{Pich:2009sp, Jung:2010ik} or the THDM with general flavour structure.
\begin{figure}[H]
\begin{center}
\includegraphics[width=8.5cm]{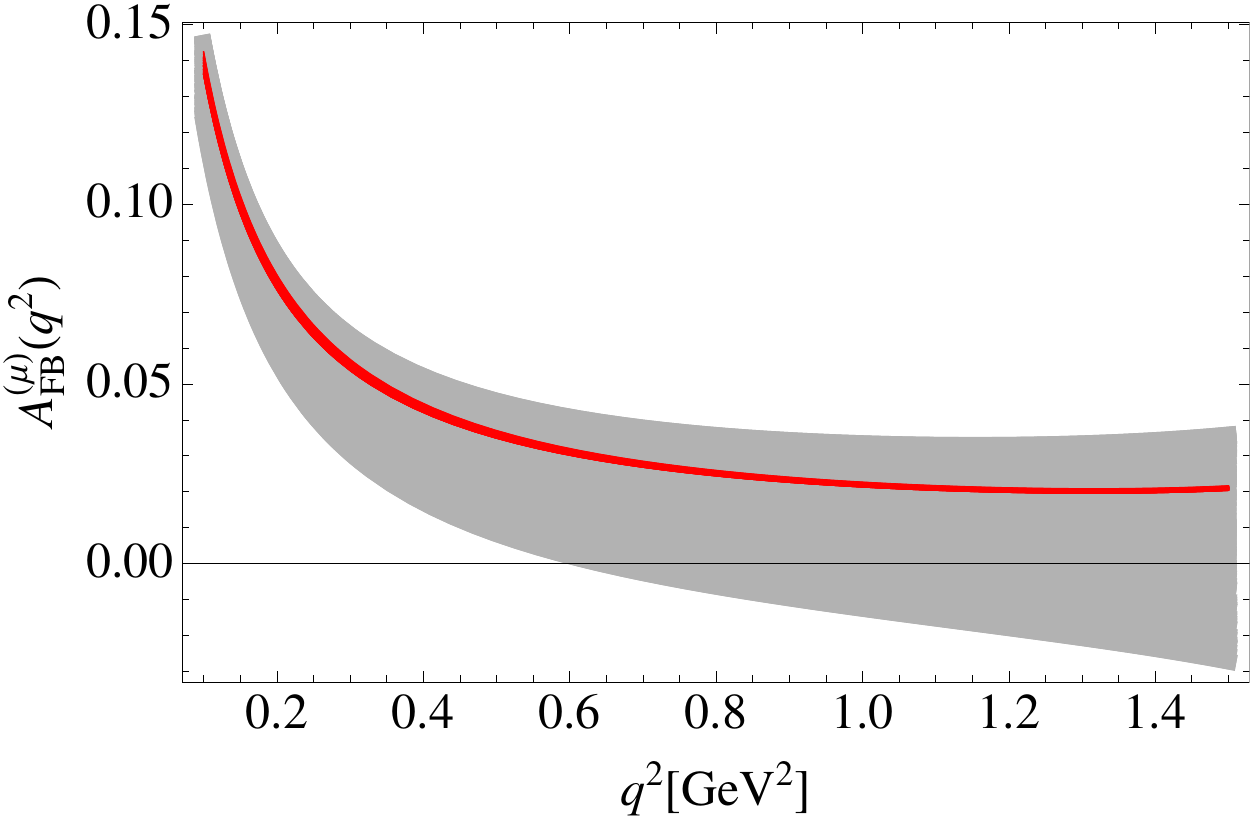}\label{Fig.2}
\caption{Comparison of the shape of forward-backward asymmetry $A_{FB}^{(\mu)}(q^2)$ in the SM (red) with the deviations (grey) induced by currently allowed values of $c_S^{(\mu)}$ couplings. Coloured bands represent the form factor uncertainties.}
\end{center}
\end{figure}
\subsection{NP in transversal muon polarization}
The relative complex phase between the non-standard scalar Wilson coefficient and the $V_{cs}$ element of the CKM matrix is a possible new source of the CP violation. The total decay rate does not offer an independent information about such effects. One could measure the T-odd transverse polarization of the final charged lepton in the semileptonic D meson decays~\cite{Barranco:2013tba, Barranco:2014bva}. It follows from the CPT invariance that this observable is also CP-odd. Since its value is expected to be vanishingly small in the SM, the measured non-vanishing value would be clear sign of the NP. This observable was first theoretically introduced and experimentally studied in semileptonic K meson decays, see~\cite{Leurer:1988au, Garisto:1991fc, Abe:2006de}. The transversal polarization of the $\tau$ lepton in the semitauonic B decays has also been theoretically considered as a possible test of the beyond SM CP violating effects, see~\cite{Garisto:1994vz,Wu:1997uaa}. In the case of process with the electron in the final state, this observable remains insensitive to the corresponding scalar Wilson coefficient.
We define the transversal polarization of the muon in the process $D^+\to K^0\mu^+\nu$ as the ratio:
\begin{equation}
P_\perp^{(\mu)}=\frac{\vert \mathcal A(\vec{s})\vert^2-\vert\mathcal A(-\vec{s})\vert^2}{\vert \mathcal A(\vec{s})\vert^2+\vert\mathcal A(-\vec{s})\vert^2},\label{transversal}
\end{equation} 
where $\vec{s}\equiv (\vec{p}_K\times\vec{p}_\ell)/\vert\vec{p}_K\times\vec{p}_\ell\vert$ denotes the unit vector perpendicular to the $K\ell$ decay plane and $A(\pm\vec{s})$ is the amplitude for spin projections along $\vec{s}$. The small value of $P^\perp_{(\ell)}$ is in the SM generated by the final state interactions. For example, the electromagnetic effects produce the value of the order $10^{-6}$ in the process $K^+\to\pi^0\mu^+\nu$~\cite{Ginsberg:1974vr}. The theoretical computations of the contributions of the final state interactions on this observable in the semileptonic D decays is currently lacking, but we expect that it is small enough that it can be neglected. 
The contribution to the numerator of~\eqref{transversal} arises from the interference between the SM and the scalar amplitudes~\cite{Leurer:1988au, Garisto:1991fc, Wu:1997uaa, Garisto:1994vz}, namely
\begin{equation}
P_\perp^{(\mu)}(q^2,E_\mu)=\bigg(\frac{d\Gamma}{dq^2 dE_\mu}\bigg)^{-1}\,\kappa(q^2,E_\mu)\operatorname{Im}\big(h_0(q^2)h_t^{\ast}(q^2)\big).
\end{equation}
The NP contribution is encoded in the modification of the helicity amplitude $h_t(q^2)$ (see Eq.~\eqref{eq:1}). The function $\kappa(q^2,E_\mu)$ is given by
\begin{equation}
\kappa(q^2,E_\mu)=-2\sqrt{\frac{r_\mu}{\lambda}}\bigg[\bigg(\frac{4E_\mu}{m_D^2}-4r_\mu\bigg)\bigg((1-r_K-r_q)^2-4r_K\bigg)-4\bigg(-\frac{2E_\mu}{m_D}+2r_K+r_\mu+\frac{E_\mu(1-r_K-r_q)}{m_D}+r_q\bigg)^2\bigg]^{1/2},
\end{equation}
where $r_\mu=m_\mu^2/m_D^2$, $r_K=m_ K^2/m_D^2$, $r_q=q^2/m_D^2$ and $E_\mu$ is the energy of the muon in the rest frame of the decaying D meson.
The average of the transversal lepton polarization over the specific kinematic region:
\begin{equation}
\langle P_\perp^{(\mu)}\rangle =\frac{\int dq^2 dE_\mu\,P_{\perp}^{(\mu)}(q^2,E_\mu)\frac{d^2\Gamma}{dq^2dE_\mu}}{\int dq^2 dE_\mu\frac{d^2\Gamma}{dq^2dE_\mu}},
\end{equation}
yields the quantity that is the measure of the difference between the number of charged leptons with their spins pointing above and below the decay plane, divided by their total number. While in the SM the value of $\langle P_\ell^\perp\rangle$ is expected to be very small (close to zero), for the presently allowed values $c_S^{(\mu)}\simeq\pm\, 0.1\,i$ we find the maximally allowed value $\langle P_\perp^{(\mu)}\rangle\simeq \pm 0.2$.

\section{Right handed current}
\label{Right handed section}
We now study the constrains on the effective operator that involves the right-handed current $\bar{s}\gamma_\mu P_R c$. The Wilson coefficient is expected to be of the form of a product of the universal coupling~$\epsilon_R$ and the corresponding quark mixing matrix element in the right-handed quark sector, see e.g.~\cite{Buras:2010pz}. In the past few years the right handed quark currents have been studied as a possibility to accommodate the tensions between the values of the $\vert V_{ub}\vert$ extracted from the exclusive and inclusive (semi)leptonic decays~\cite{Crivellin:2009sd, Buras:2010pz}. 
The right-handed current would modify the extraction of the $\vert V_{cs}\vert$ in the following way
\begin{equation}
\vert V_{cs}(1+c_{V,R})\vert =\vert V_{cs}(D\to K\ell\nu)\vert_{SM/exp},\quad \vert V_{cs}(1-c_{V,R})\vert =\vert V_{cs}(D_s\to\ell\nu)\vert_{SM/exp},\label{modifications}
\end{equation}
where $\vert V_{cs}\vert_{SM-exp}$ denote the values extracted from the comparison of the experimental and predicted (in the SM) values of the branching fractions.
We assume the $c_{V,R}$ to be real-valued, lepton universal and lot smaller than one, so that the above relations can be expanded to first order in this coefficient. Using the values $\vert V_{cs}(D\to \ell\nu)\vert_{SM/exp}=1.010(20)$, from Ref.~\cite{Bazavov:2014wgs}, and $\vert V_{cs}(D\to K\ell\nu)\vert_{SM/exp}=0.963(15)$, from Ref.~\cite{Koponen:2013tua}, we obtain the limits:
\begin{equation}
\vert V_{cs} \vert =0.987\pm 0.013, \quad c_{V,R}=-0.023\pm 0.013.\label{constraint-gR}
\end{equation}
The resulting value of $c_{V,R}$ coupling is compatible with zero at the $95\%$ C.L., while the value of $\vert V_{cs}\vert$ is compatible with the result of the global unitarity fit~\cite{Charles:2004jd}. 

The $c_{V,R}$ can be further constrained in $D\to V\ell\nu$ decay modes. The HPQCD Colalboration recently calculated the ratio of the form factors $V(0)/A_1(0)=1.72(21)$ for the process $D_s\to \phi e\nu_e$~\cite{Donald:2013pea}. 
This ratio is modified by the presence of the right handed currents via:
\begin{equation}
V(0) \rightarrow (1+c_{V,R}) V(0),\quad A_1(0)\rightarrow (1-c_{V,R}) A_1(0).
\end{equation}
Comparison of the lattice result with the value measured by the BaBar Collaboration $V(0)/A_1(0)=1.849\pm 0.11$~\cite{Aubert:2008rs} results in interval
\begin{equation}
-0.03\leq c_{V,R}\leq 0.1\,.
\end{equation}
Once the lattice results in these processes are further refined, the more detailed constraints on the right-handed contributions could be performed with the use of the angular distributions, as explained in the Ref.~\cite{Bernlochner:2014ova}.
\section{Conclusions}
We have investigated leptonic and semileptonic $c\to s \ell \bar{\nu}_\ell$ transition of charm mesons using the effective Lagrangian approach. The most constraining processes for the pseudoscalar couplings are leptonic decays, due to the very good knowledge of the $D_s$ meson decay constant obtained by the lattice QCD and the latest precise measurements. 
The branching ratios for the decay $D \to K^\ast\ell\nu$, the ratio of the decay widths for the longitudinally and transversely polarised $K^\ast$ have already been measured. We use the existing experimental result to look for an additional constraint on the pseudoscalar coupling. In order to obtain better bound one should have precise lattice determination of $A_0(q^2)$ form factor as well as more precise experimental results.

The scalar Wilson coefficients can be constrained from $D \to K \ell \nu$ decay modes. The most interesting observables in this respect are the forward-backward asymmetry and the CP-violating transverse muon polarization in the decay involving muons in the final state. The deviations from the SM in these observables are currently allowed. We found out that the ratio $R_{\mu/e}(q^2)\equiv \frac{d\Gamma^{(\mu)}}{dq^2}/\frac{d\Gamma^{(e)}}{dq^2}$ might be used to test lepton flavour violation. By allowing the first generation of leptons to interact as in the SM, and new physics to affect the second generation, we find that this ratio is currently allowed to deviate from the SM value by $10-20\%$, depending on the $q^2$.
Finally, we constrain the Wilson coefficient of the right handed current in the charm Cabibbo allowed (semi)leptonic precesses using both experimental results on $D_s\to \ell\nu$ and the lattice QCD calculation for the form factors ratio in $D_s\to \phi e\nu_e$. The both constraints are compatible. The future experiments on charm meson leptonic and semileptonic decays as well as lattice QCD studies will lead to very strong constraints on possible NP contributions.\\

\textbf{Acknowledgements.}
I.N. is supported in part by the  Bundesministerium f\"ur Bildung und Forschung and the research of S.F. has been supported by Slovenian research agency ARRS.  

\begin{appendix}
\section{Hadronic helicity amplitudes for $D\to V\ell\nu$}\label{Appendix1}
The non-vanishing hadronic helicity amplitudes for the $P\to V\ell\nu$ decay process are given by the following formulas:
\begin{equation}
\begin{split}
H_\pm(q^2)&=\mp\frac{\sqrt{\lambda(m_P^2,m_V^2,q^2)}}{m_P+m_V}V(q^2)+(m_P+m_V)A_1(q^2)\\
H_0(q^2)&=\frac{1}{2m_V\sqrt{q^2}}\bigg[(m_P+m_V)(m_P^2-m_V^2-q^2)A_1(q^2)-\frac{\lambda(m_P^2,m_V^2,q^2)}{m_P+m_V}A_2(q^2)\bigg]\\
H_t(q^2)&=\,\bigg[1-c_P^{(\ell)}\frac{q^2}{m_\ell(m_q+m_{\bar{q}})}\bigg]\frac{\sqrt{\lambda(m_P^2,m_V^2,q^2)}}{\sqrt{q^2}}A_0(q^2).\label{helicity amplitudes}
 \end{split}
\end{equation}
\end{appendix}

\end{document}